\documentclass[twocolumn,superscriptaddress,showpacs,preprintnumbers,amsmath,amssymb]{revtex4}

\usepackage{graphicx}
\usepackage{dcolumn}
\usepackage{bm}
\usepackage{rotating}

\begin{document}

\preprint{0705.0843}

\title{Simulation via Direct Computation of Partition Functions}

\author{Cheng~Zhang}
\affiliation{Department of Bioengineering, Rice University,
Houston, Texas 77005, USA}

\author{Jianpeng~Ma}
\email{jpma@bcm.tmc.edu}
\affiliation{Department of Bioengineering,
Rice University, Houston, Texas 77005, USA}
\affiliation{ Verna and Marrs McLean Department of Biochemistry
and Molecular Biology, Baylor College of Medicine, One Baylor
Plaza, BCM-125, Houston, Texas 77030, USA }

\date{\today}

\begin{abstract}
In this paper, we demonstrate the efficiency of simulations via
direct computation of the partition function under various
macroscopic conditions, such as different temperatures or volumes.
The method can compute partition functions by flattening histograms,
through the Wang-Landau recursive scheme, outside the energy space.
This method offers a more general and flexible framework for
handling various types of ensembles, especially the ones in which
computation of the density of states is not convenient. It can be
easily scaled to large systems, and it is flexible in incorporating
Monte Carlo cluster algorithms or molecular dynamics. High
efficiency is shown in simulating large Ising models, in finding
ground states of simple protein models, and in studying the
liquid-vapor phase transition of a simple fluid. The method is very
simple to implement and we expect it to be efficient in studying
complex systems with rugged energy landscapes, e.g., biological
macromolecules.
\end{abstract}


\maketitle

In recent years, methods for Monte Carlo (MC) simulation have been
dramatically improved over the traditional Metropolis
algorithm~\cite{metropolis}.  A large class of MC methods are those
based on the flat energy histogram, such as the multicanonical
ensemble method~\cite{muca}, the entropic sampling
method~\cite{entropic}, the density of states (DOS)
method~\cite{wl}, and the statistical temperature method~\cite{st}.
In this study, we demonstrate the efficiency of an alternative
sampling method, which simultaneously and directly computes the
partition function at various values of a certain macroscopic
variable, e.g., temperature $T$ or volume $V$.  Since one does not
know the partition function in advance, the partition function at
different values of a chosen variable is initially set to unity and
continuously modified throughout the simulation until convergence.

We first demonstrate the case of sampling based on a number of
discrete values of temperature.  In this case, a number of sampling
temperatures are set over the temperature range of interest. Similar
to the expanded ensemble method or the simulated tempering
method~\cite{expanded}, two types of MC moves are used: an energy
move under a fixed temperature and a temperature move under a fixed
energy. Before each MC step, a fixed probability is used to
determine which type of move the system takes. For the energy move,
the Metropolis algorithm is performed at the present (reciprocal)
temperature $\beta$.  For the temperature move, another temperature
$\beta'$ is randomly chosen, and the following acceptance
probability is used to accept the move:
\begin{equation}
    \mbox{Acc}(\beta\rightarrow\beta') =
    \min\left\{
    1,
    \frac{\exp(-\beta'E)/\tilde Z_{\beta'}}{\exp(-\beta E)/\tilde Z_\beta}
    \right\}.
\label{eq:acc}
\end{equation}
Here $E$ is the present energy;  $\tilde Z_\beta$ and $\tilde
Z_{\beta'}$ are the values of the estimated partition function at
temperatures $\beta$ and $\beta'$, respectively. The partition
function is ``estimated'' because it is unknown in advance. After
each MC step, the estimated partition function at the present
temperature is multiplied by a factor $f>1$~\cite{wl}.  This can be
written as,
\begin{equation}
   \ln \tilde Z_\beta \rightarrow \ln \tilde Z_\beta + \ln f.
\label{eq:update}
\end{equation}

Similar to the WL algorithm,  it is shown that by repeating the
above procedure for a fixed $f$, the estimated partition function
can eventually converge within certain fluctuations proportional to
$\sqrt{\ln f}$~\cite{wlproof,Z}. Moreover, due to the frequently
modified acceptance probability, the additional errors in the
estimated partition function (due to violation of the detailed
balance condition) are larger in a stage with a larger $\ln f$.
Therefore, the value of $\ln f$ should be gradually decreased to
improve the accuracy of the estimated partition function.  In
practice, the whole simulation is separated into several stages,
each marked by a different value of $\ln f$~\cite{wl}. In passing
from one stage to the next, $\ln f$ is modified to $(\ln f)
/n$~\cite{wl}.  We use $n=\sqrt{10}$ in this study so that $\ln f$
is decreased by an order of magnitude every two stages (the
procedure for optimizing the $\ln f$ of each intermediate stage will
be given in a forthcoming paper~\cite{Z}).  At the end of the
simulation, $\ln f$ is reduced to a tiny number such that violation
of the detailed balance condition is negligible. For each $f$ stage,
if the simulation runs for sufficient number of steps, each
temperature receives on average an equal number of visits, i.e., a
flat temperature histogram is achieved.  Here the term ``temperature
histogram'' refers to the number of visits to each discrete
temperature instead of to a temperature interval.  The simulation is
allowed to enter the next $f$ stage when the histogram fluctuation
falls below a cutoff percentage~\cite{wl}.

An alternative approach is to fix the number of simulation steps by
$C/\sqrt{\ln f}$ for an $f$ stage. It can be shown that the two
approaches are equivalent for sufficiently long
simulations~\cite{Z}. The constant $C$ can be estimated from a few
initial $f$ stages. The second approach ensures a better convergence
for a stage with a smaller $\ln f$.

In principle, any set of sampling temperatures of interest can be
used.  However, two consecutive temperatures must be close enough to
allow sufficiently frequent temperature transitions.  This requires
a certain overlap between the energy distributions of two
neighboring temperatures.  This condition can be expressed as
$\Delta T \sim \sqrt{\langle \Delta E^2\rangle}/C_V \sim
T/\sqrt{C_V}$ , where $C_V$ and $\sqrt{\langle\Delta E^2\rangle}$
are the heat capacity and energy fluctuation at temperature $T$,
respectively. Therefore, the number of sampling temperatures is
roughly proportional to $\sqrt N$ (except around the critical
region), where $N$ is the system size. This feature is advantageous
for larger systems, which is also a merit of the parallel tempering
method~\cite{replica}, but the latter does not deliver the partition
function quickly.

\begin{figure}[h]
  \begin{minipage}{\linewidth}
    \begin{center}
        \includegraphics[angle=-90,width= \linewidth]{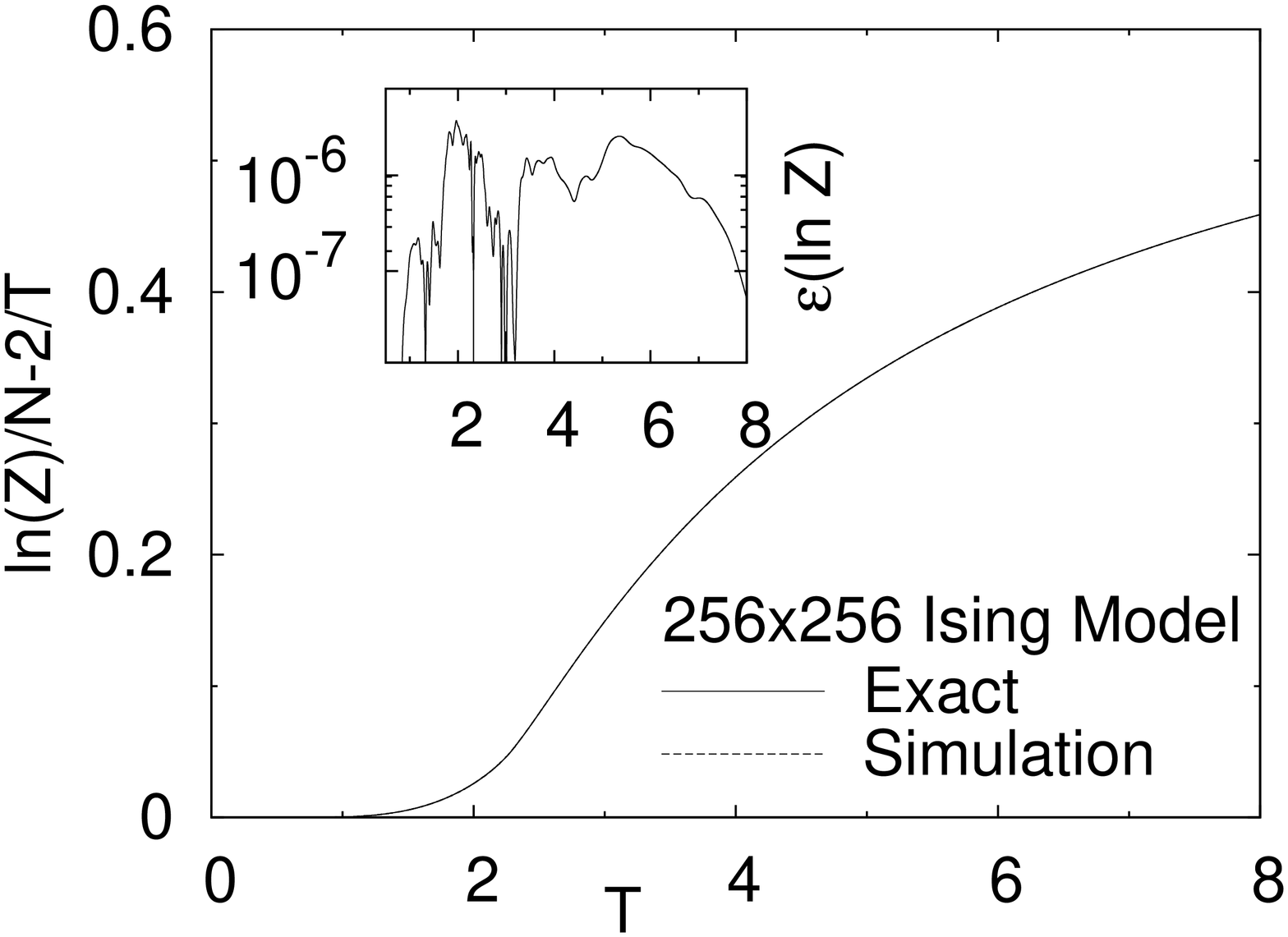}
    \end{center}
  \end{minipage}%
\\
  \begin{minipage}{\linewidth}
    \begin{center}
        \includegraphics[angle=-90,width= \linewidth]{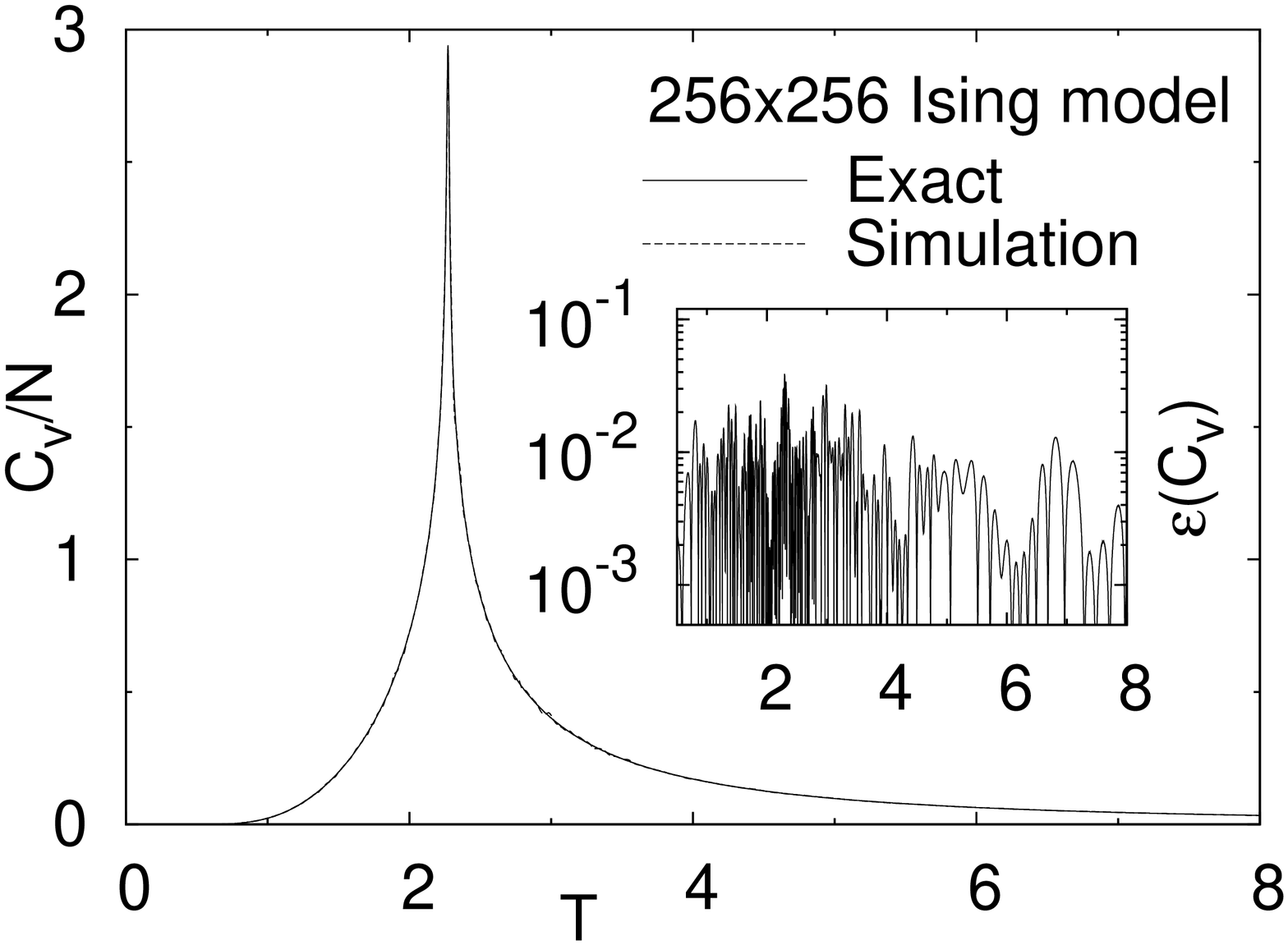}
    \end{center}
  \end{minipage}%
  \caption{\label{fig:256}
Results for the $256\times 256$ Ising model. The upper panel shows
the partition function as a function of temperature.  The curve is
shown for $\ln Z$ per spin with the contribution of the two ground
states subtracted. The lower panel shows the heat capacity per spin
as a function of temperature.  The relative errors are shown in the
insets for both panels.  }
\end{figure}

The algorithm was first tested on the $256\times256$ square lattice
Ising model. A wide temperature range, $T \in [0,8]$, was simulated
in a single simulation. Since the sampling temperature increment of
an efficient simulation should be inversely related to the heat
capacity as discussed above (nonuniform temperature setup is known
to be advantageous~\cite{nutemp}), for this large system, sampling
temperatures were distributed based on the roughly estimated heat
capacity (e.g., that from simulation of a smaller system).
Accordingly, the entire temperature range was partitioned into 13
subranges. Sampling temperatures were linearly distributed inside
each subrange with a different increment. The temperature subranges
and their increments were $(0.1, 1.0 | 0.1)$, $(1.0, 1.8 | 0.04)$,
$(1.8, 2.0 | 0.02)$, $(2.0, 2.2 | 0.005)$, $(2.2, 2.25 | 0.0025)$,
$(2.25, 2.3 | 0.002)$, $(2.3, 2.35 | 0.005)$, $(2.35, 2.5 | 0.01)$,
$(2.5, 2.7 | 0.02)$, $(2.7, 3.6 | 0.05)$, $(3.6, 5.0 | 0.07)$,
$(5.0, 6.0 | 0.1)$, and $(6.0, 8.0 | 0.2)$. Here the notation for
each subrange is (beginning temperature, ending temperature $|$
increment).  In total, there were 218 sampling temperatures. Each
time the probability of choosing temperature over energy moves was
0.1\% (this number should be larger for smaller systems).  The
modification factor $\ln f$ was decreased from 1.0 to $10^{-9}$, the
number of MC steps for stage $f$ was $100/\sqrt{\ln f}$ sweeps, so
the whole simulation took $7.2\times10^6$ sweeps. Thermodynamic
quantities at temperatures other than the sampled temperatures can
be calculated using the multiple histogram method~\cite{mhistogram}.
Histograms from the last $f$ stage were used. The exact results of
the Ising model were also calculated using the method by Ferdinand
and Fisher~\cite{isingexact}. The relative errors of the partition
function, energy, entropy, and heat capacity were no larger than
0.00064\%, 0.071\%, 1.1\%, and 3.9\%, respectively.
Fig.~\ref{fig:256} shows the results for the partition function and
heat capacity. For comparison, the WL algorithm was applied to the
same system using 15 independent simulations, and the maximum
relative errors of the free energy, energy, entropy, and heat
capacity were 0.0008\%, 0.09\%, 1.2\%, and 4.5\%,
respectively~\cite{wl}. The simulation cost of the WL algorithm  was
$6.1\times10^6$ sweeps~\cite{wl}. However, the acceptance
probabilities for energy moves can be precalculated to avoid
expensive exponential computation in our case. The above simulation
was finished in 10 hours on a single Intel Xeon processor~(2.8~GHz).

Next, we introduce a variation of the above algorithm that tries to
find the transition temperature automatically and to spend more
effort sampling around that. This feature is desirable if the
transition temperature is not roughly estimated in advance.  This
can be achieved by using a modified updating scheme, to let the
system visit each temperature with a different frequency $w_\beta$.
In the acceptance probability Eq.~(\ref{eq:acc}), the values,
$\tilde Z_\beta$ and $\tilde Z_{\beta'}$, of the estimated partition
function are replaced by $\tilde Z_\beta/w_\beta$ and $\tilde
Z_{\beta'}/w_{\beta'}$, respectively, whereas the updating scheme
Eq.~(\ref{eq:update}) is changed to $\ln \tilde Z_\beta \rightarrow
\ln \tilde Z_\beta +\ln f /w_\beta$. The temperature histogram is
constructed in such a way that the total number of visits to a
particular temperature $\beta$ is now divided by its associated
frequency $w_\beta$. To focus sampling around the transition
temperature, the frequency $w_\beta$ can be specified as an
increasing function of the heat capacity. Since the values of the
heat capacity are unknown in advance, they are updated at the end of
each $f$ stage and are used in the next stage.  The modified
algorithm was tested on the same $256\times256$ Ising system.  The
frequency $w_\beta$ at temperature $\beta$ was set as the square of
the heat capacity per spin. Sampling temperatures were uniformly
distributed over the whole range, $T\in[0,8]$, with a fixed
increment $\Delta T=0.002$. The probability of choosing temperature
over energy moves was raised to 10\%. The value of $\ln f$ was
lowered from 1.0 to $\sqrt{10} \times 10^{-9}$. The simulation was
kept running at each $f$ stage until the temperature histogram
fluctuation was lowered below 50\%. The last stage was purposely
extended to $5.0\times10^6$ MC sweeps to accumulate more statistical
data. Totally, $9.8\times 10^6$ sweeps were used. The relative
errors of the free energy, the energy, and the heat capacity were no
larger than 0.000~45\%, 0.055\%, and 4.0\%, respectively.

It is also possible to realize rejection-free, hence more efficient,
temperature transitions. First, the relative probability at each
temperature  $\beta_i$, $P_i = \exp(-\beta_i E)/\tilde Z_{\beta_i}$,
is calculated for the present energy $E$.  Next, the accumulated
probability for each temperature, $Q_i=\sum_{j \le i} P_j/\sum_j
P_j$, is also calculated, to form a series of brackets, $[Q_{i-1},
Q_i)$, $i=1,2,\ldots$, with $Q_0=0$. If a uniform random number $r
\in [0,1)$ falls in the $i$th bracket, $\beta_i$ will be chosen as
the next temperature. This type of temperature move is analogous to
the heat bath algorithm for energy moves~\cite{heatbath}. It is
relatively expensive because of many exponential calculations.
However, this expense is negligible if a more expensive
non-Metropolis algorithm is used for the energy move.
As an example, the Swendsen-Wang cluster algorithm~\cite{cluster}
was used as the energy move on large two-dimensional Ising models.
To improve the efficiency, the energy and temperature moves were
merged in such a way that each energy move was immediately followed
by a rejection-free temperature move. Simulations were performed on
critical temperature windows estimated by $|T-T_c| \sim L^{-\nu}$.
Here $\nu=1$ is the critical exponent, and $T_c$ is the critical
temperature. About 10$-$20 sampling temperatures were distributed in
each window. Parameters and results are listed in
Table~\ref{tab:cluster}. The efficiency is clear in terms of the
number of simulation steps required to reach the desired accuracy.

\begin{table}[h]
\caption{\label{tab:cluster} Results for $L\times L$ Ising models
using the Swendsen-Wang cluster algorithm \cite{cluster} as the
energy move. Maximum relative errors were calculated by assuming the
errors at the left boundary to be zeros. Here, $T_-$ and $T_+$
define the temperature window, and $\Delta T$ defines the increment.
}
\begin{ruledtabular}
\begin{tabular}{lccccccc}
$L$  & $(T_-, T_+ | \Delta T)$ & MC steps & $\epsilon(\ln Z)$  & $\epsilon(C_V)$ \\
\hline
64  & (2.0, 2.9 $|$ 0.1) & $0.7\times10^6$ & $4.0\times 10^{-6}$ & 1.6\% \\
128 & (2.1,2.6 $|$ 0.05)  & $2.0\times10^6$ & $1.2\times 10^{-6}$ & 1.1\% \\
256 & (2.2,2.42 $|$ 0.02)  & $2.9\times10^6$ & $3.6\times 10^{-7}$ & 1.4\%  \\
512 & (2.2,2.34 $|$ 0.01)  & $3.1\times10^6$ & $1.0\times 10^{-7}$ & 1.0\%  \\
1024 &(2.24,2.30$|$0.005)  & $3.1\times10^6$ & $6.9\times 10^{-8}$ & 1.4\%  \\
\end{tabular}
\end{ruledtabular}
\end{table}

\begin{figure}[h]
  \begin{minipage}{.5\linewidth}
    \begin{center}
        \includegraphics[angle=-180,width= 1.2 \linewidth]{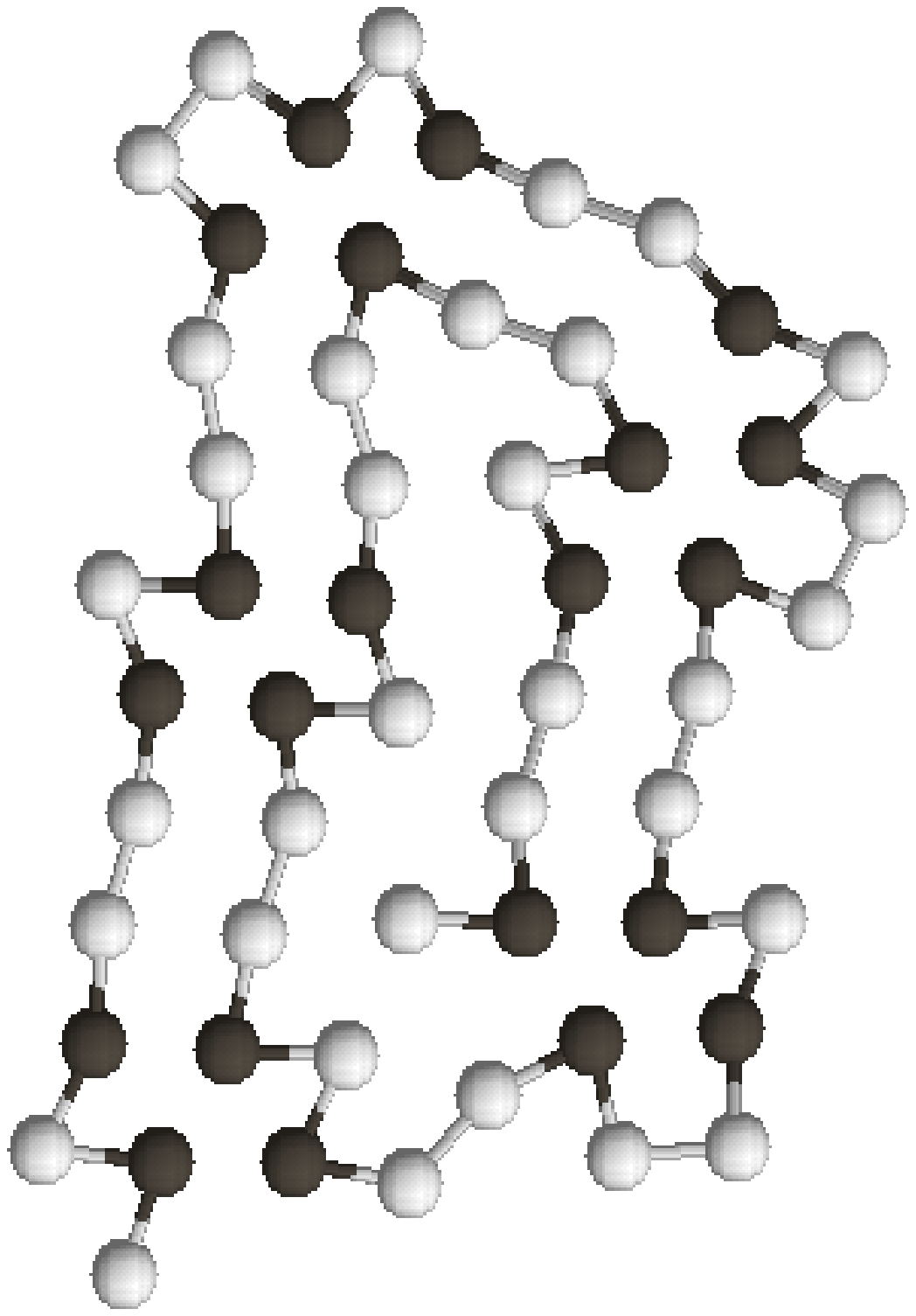}
    (a)
    \end{center}
  \end{minipage}%
  \begin{minipage}{.5\linewidth}
    \begin{center}
        \includegraphics[angle=-90,width= 1.2 \linewidth]{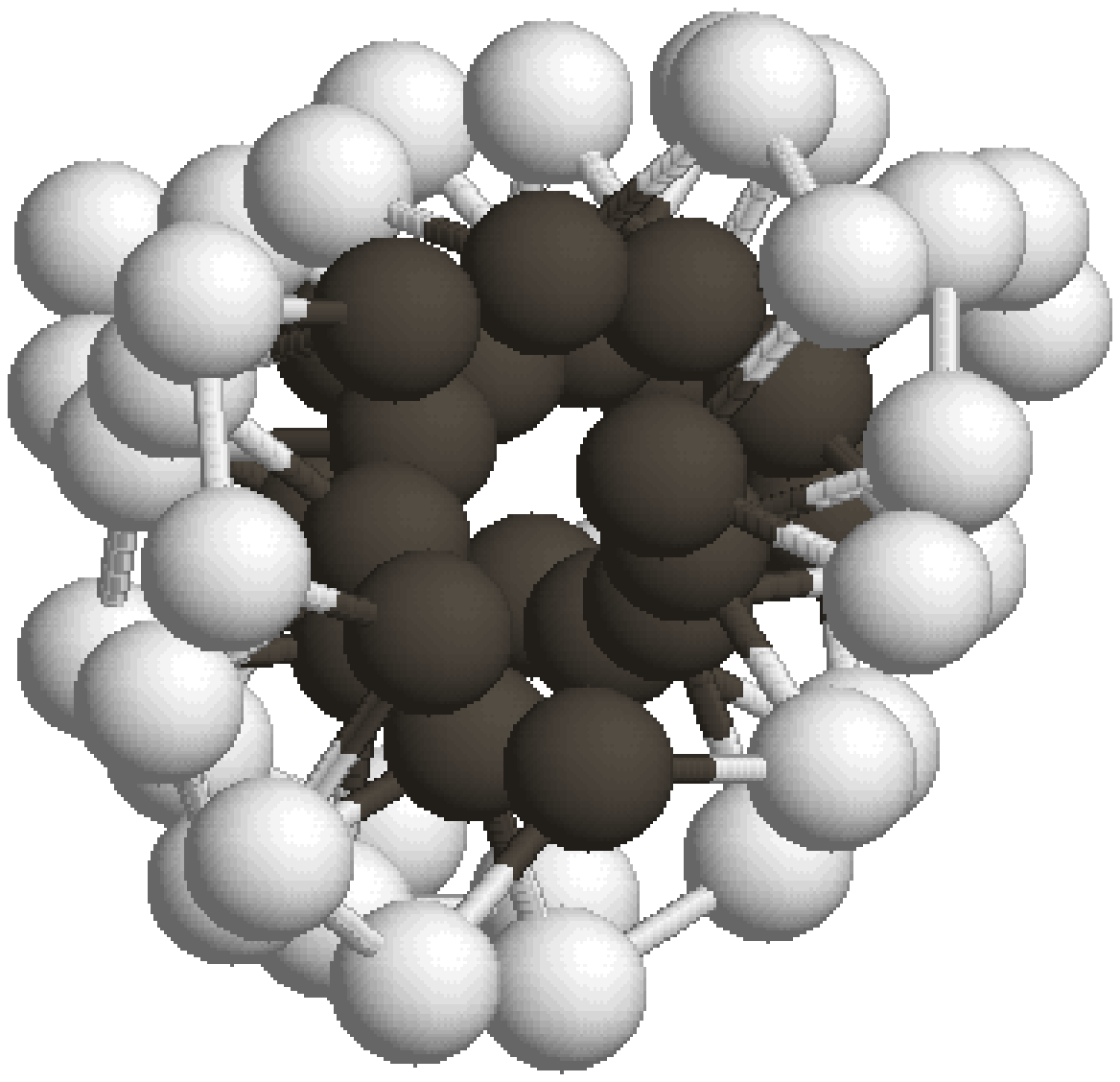}
    (b)
    \end{center}
  \end{minipage}%
\\
  \begin{minipage}{.5\linewidth}
    \begin{center}
        \includegraphics[angle=-90,width= 1.2 \linewidth]{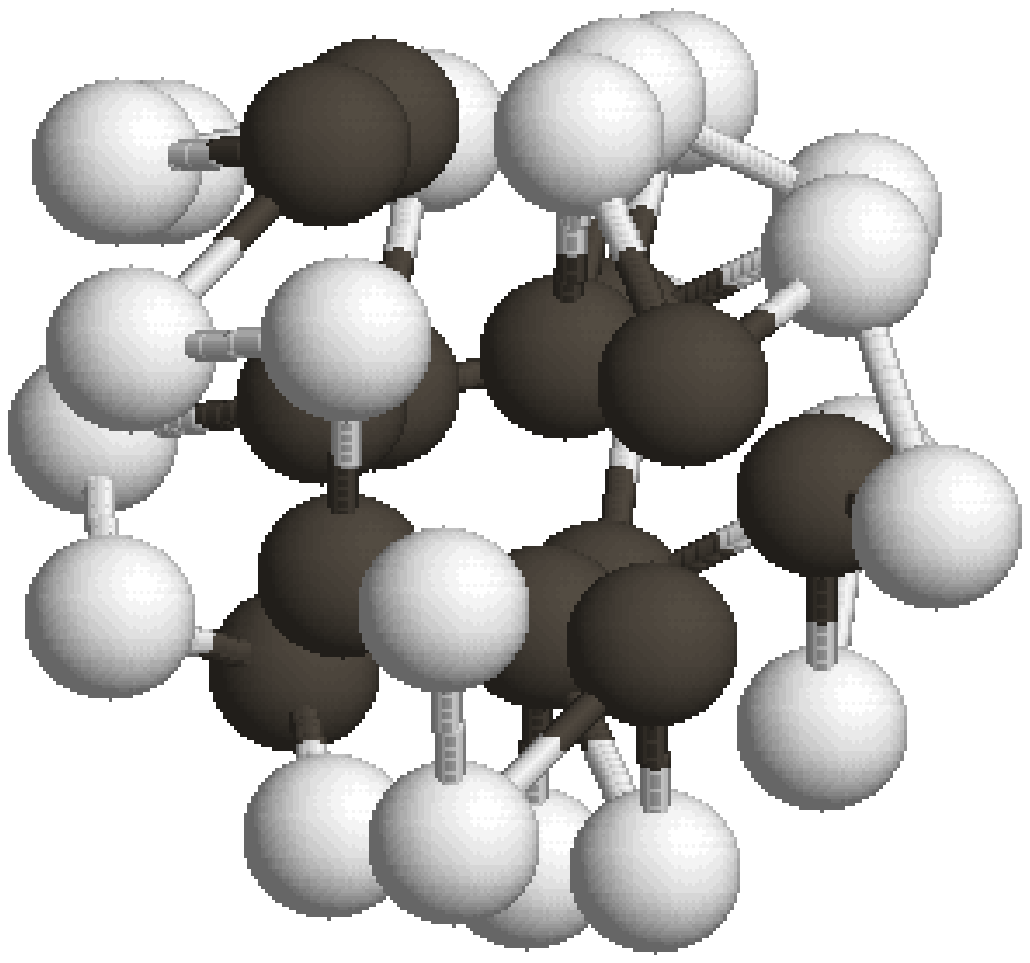}
    (c)
    \end{center}
  \end{minipage}%
  \begin{minipage}{.5\linewidth}
    \begin{center}
        \includegraphics[angle=-90,width= 1.2 \linewidth]{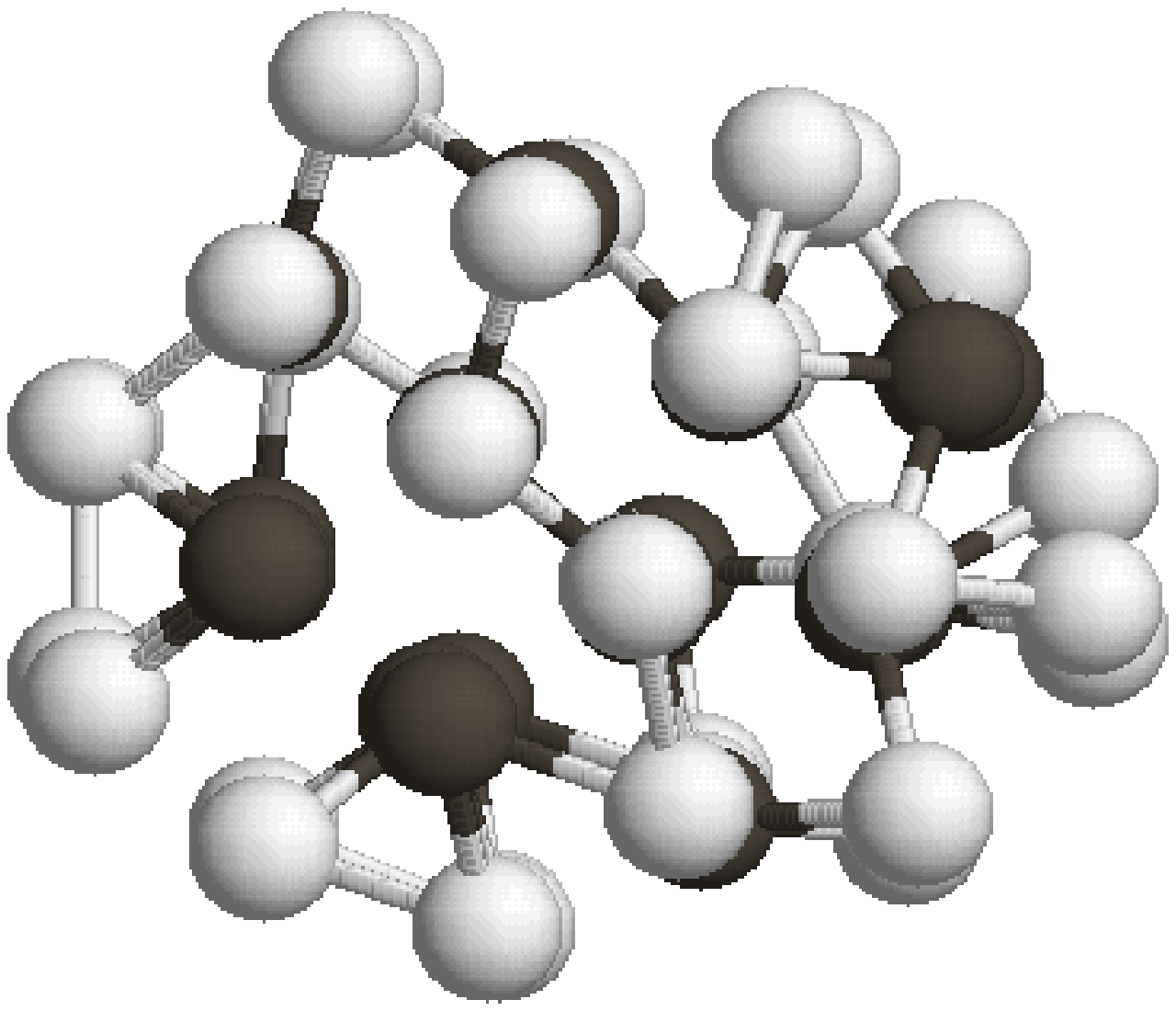}
    (d)
    \end{center}
  \end{minipage}
\caption{\label{fig:ab} Lowest-energy configurations of $AB$
proteins (black, $A$; white, $B$).
(a)~2D, 55mer, model~I.
(b)~3D, 55mer, model~I.
(c)~3D, 34mer, model~II.
(d)~3D, 55mer, model~II. }
\end{figure}

Molecular dynamics (MD) can be used as an energy move as well. In
this case, the probability of taking temperature over energy moves
is 50\%. Constant-temperature MD (a length-5 Nos\'e-Hoover
chain~\cite{nhchain} with force-scaling~\cite{forcescaling}) is used
as a (potential-)energy move~\cite{st}.  The thermostat temperature
$T_0$ was set to be $0.5$. The simulations were used to find ground
states of AB protein models~\cite{ab}. We were able to find all
known ground states~\cite{acmc,elp,csa, st}, and several new ones
with lower energies. Table~\ref{tab:ab} lists the new ground-state
energies, and Fig.~\ref{fig:ab} shows the corresponding
configurations. Comparing our results (for model I~\cite{ab}) with
those from the statistical temperature method~\cite{st}, the new
ground state of the two-dimensional (2D) 55mer,
Fig.~\ref{fig:ab}(a), has a different topology in the two inner
strands; the new ground state of the three-dimensional (3D) 55mer,
Fig.~\ref{fig:ab}(c), has a more compact configuration. In both
cases, our ground states have black-black clusters (strong
attractions) that are more favorably packed with no exposed black
beads.

\begin{table*}
\caption{\label{tab:ab} Lowest energies of $AB$ proteins with
Fibonacci sequences. Results are compared with those from the
annealing contour Monte Carlo (ACMC)~\cite{acmc}, the energy
landscape paving (ELP)~\cite{elp}, the conformational space
annealing (CSA)~\cite{csa}, and the statistical temperature
molecular dynamics (STMD)~\cite{st}.}
\begin{ruledtabular}
\begin{tabular}{lcccccc}
protein         &   ACMC &ELP & CSA & STMD & This work\\
\hline
2D, 55mer, model I   & $-18.7407$ &            &  $-18.9110$ & $-18.9202$ & $-19.2570$  \\
3D, 55mer, model I   &            &  $-42.438$ & $-42.3418$  & $-42.5789$ & $-44.8765$  \\
3D, 34mer, model II  & $-94.0431$ & $-92.746$  & $-97.7321$  &            & $-98.3571$  \\
3D, 55mer, model II  &$-154.5050$ & $-172.696$ & $-173.9803$ &            & $-178.1339$ \\
\end{tabular}
\end{ruledtabular}
\end{table*}

\begin{figure}[h]
  \begin{minipage}{ \linewidth}
    \begin{center}
        \includegraphics[angle=-90,width=  \linewidth]{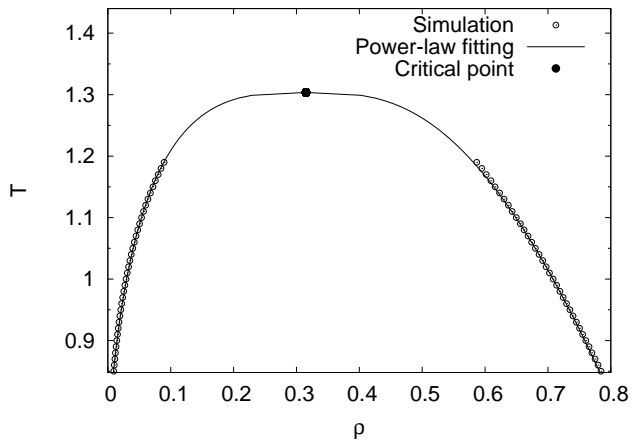}
    \end{center}
  \end{minipage}%
  \caption{\label{fig:phase} Phase diagram for the 108-particle Lennard-Jones system.
  The empty circles are results of simulations,
  the solid line is from power-law fitting,
  and the solid circle represents the estimated critical point for this small system.}
\end{figure}

The WL-type algorithms have also been applied to Lennard-Jones
simple liquid systems~\cite{wllj} through computing the
multidimensional DOS. Here, we demonstrate that the simulation can
be carried out using volume, instead of temperature, as the sampling
variable, where the temperature and particle number are held
constant. Each volume move can be implemented as a change of the
scale of the system. Therefore, it is convenient to adopt reduced
coordinates $\mathbf s=\mathbf r/\sqrt[3]{V}$. The partition
function is factorized to the ideal gas part $Z_{ig}$, and a
potential part $Z_V$, i.e., $Z=Z_{ig} Z_V$, where $Z_V \equiv
(1/V^N) \int d \mathbf r^N \exp[-\beta U (\mathbf r^N)] = \int d
\mathbf s^N \exp[-\beta U (\mathbf s^N; V)]$. Thus, we can
dynamically compute the potential part of the partition function
$Z_V$, instead of $Z$, in the acceptance probability
Eq.~(\ref{eq:acc}).
This method was used to study the liquid-vapor transition of a
108-particle Lennard-Jones system with half-box truncation and
periodic boundary conditions. After the simulation, the Helmholtz
free energy can be obtained through $F=F_{ig}- \ln Z_V / \beta$, and
the Gibbs free energy profile under pressure $p$ can be derived
through $G=F+pV$, at each sampling volume (or density). For each
simulation under a fixed temperature, the transition pressure was
first determined by equalizing the two minima on the Gibbs free
energy curve; the values of liquid density $\rho_+$ and vapor
density $\rho_-$ were also determined correspondingly.  Simulations
were performed under different temperatures $T\in[0.85, 1.20]$, with
increment $\Delta T=0.01$. To accurately determine the position of
coexistence densities, the sampling density increments $\Delta \rho$
were 0.002 and 0.0005 around the roughly estimated liquid and vapor
coexistence densities, respectively, whereas the transition region
was filled by a larger increment $\Delta \rho=0.005$. Typically,
about 300 volume sampling points were used in a single simulation.
The computed vapor-liquid coexistence curve is shown in
Fig.~\ref{fig:phase}. The relation $\rho_{\pm} -\rho_c \sim a |T_c -
T| \pm b |T_c - T|^\beta$ (the critical exponent
$\beta=0.3258$~\cite{beta}) was used to extrapolate the critical
temperature $T_c$ and the critical density $\rho_c$  based on the
corresponding power-law regions. The estimated critical temperature
$T_c$ and critical density $\rho_c$ were 1.304 and 0.315,
respectively. The results for this small system are consistent with
those of the infinite system (e.g., $T_c=1.3123$ and
$\rho_c=0.3174$~\cite{lj}).

In summary, we have demonstrated the efficiency of simulations via
direct computation of the partition function.  The method has a
range of advantages.  An important one is in the
ground-state-oriented applications, such as in the protein folding
problem, in which case the WL algorithm suffers from lack of
efficient sampling around the ground state.  This is because the
location of the ground state, and hence the proper energy range over
which the sampling should be performed, is not known in advance. The
efficiency of the WL algorithm will be further reduced if the energy
landscape in the last energy bin (near the ground state) is
continuous and rugged~\cite{weakness}.  By contrast, sampling in the
temperature space does not require a priori information about the
ground state and can sample the vicinity of the ground state with
desired accuracy.

Our method can be viewed as a generalization of the DOS-based WL
algorithm~\cite{wl} since the DOS is indeed the partition function
of the microcanonical ensemble.  In the case of canonical versus
microcanonical ensembles, for example, the partition functions of
them are related by an expression, $Z(N,V,T) = \int_0^{\infty}
g(N,V,E) \exp(-\beta E) dE $, where $Z(N,V,T)$ is the canonical
partition function and $g(N,V,E)$ is the density of states or
microcanonical partition function. It is easy to see that, in the
canonical ensemble, one can fix any pair of thermodynamic parameters
and change the third one for sampling, while in the microcanonical
ensemble, it is hard to do so, e.g., one cannot fix $N$ and $E$ to
change $V$.  This indicates that there are inherent advantages in
performing simulations (such as flattening the histogram) outside
the energy space.  We thus expect the general framework to be more
flexible in handling other types of ensembles, especially the ones
in which computation of the DOS is not convenient.

J.M. acknowledges support from NIH Grant No. (GM067801) and a Welch
Grant No. (Q-1512).

\end{document}